\documentclass[mnsc,nonblindrev]{informs3}

\OneAndAHalfSpacedXI
\usepackage{dsfont}

\usepackage{natbib}
\bibpunct[, ]{(}{)}{,}{a}{}{,}%
\def\bibfont{\small}%

\TheoremsNumberedThrough     
\ECRepeatTheorems

\EquationsNumberedThrough    

\usepackage{mathptmx, amsmath, amssymb}

\usepackage[para, flushleft]{threeparttable}
\let\oldFootnote\footnote
\newcommand\nextToken\relax
\renewcommand\footnote[1]{%
    \oldFootnote{#1}\futurelet\nextToken\isFootnote}
\newcommand\isFootnote{%
    \ifx\footnote\nextToken\textsuperscript{,}\fi}
\usepackage{multirow}

\usepackage{tikz,makecell}

\usetikzlibrary{decorations.pathreplacing,angles,quotes}

\usepackage{booktabs}

\usepackage{setspace}
\usepackage{etoolbox}
\makeatletter 
\g@addto@macro\TPT@defaults{\linespread{0.8}\small}
\makeatother

\makeatletter
\let\normalsize\relax
\let\@currsize\normalsize
\makeatother

\usepackage{setspace}
\usepackage{etoolbox}
\makeatletter
\patchcmd{\@footnotetext}
  {\setspace@singlespace}{0.8}
  {}{}
\makeatother

\usepackage{csquotes}
\usepackage{graphicx}
\usepackage{subcaption}
\usepackage{float}

\usepackage{booktabs}
\usepackage{siunitx}
\usepackage{geometry}
\usepackage{xcolor}
\usepackage{tcolorbox}
\usepackage[T1]{fontenc} 

\usepackage{pdflscape}
\usepackage{booktabs}
\usepackage{graphicx}
\usepackage{dsfont}

\begin{document}


\tcbset{
  myhighlight/.style={
    colback=yellow!30,
    colframe=yellow!80!black,
    fontupper=\bfseries\Large,
    boxrule=1pt,
    arc=2mm,
    boxsep=4pt,
    left=6pt,
    right=6pt,
    top=6pt,
    bottom=6pt,
  }
}

\TITLE{AI-Assisted Programming Decreases the Productivity of Experienced Developers by Increasing the Technical Debt and Maintenance Burden}

\ARTICLEAUTHORS{%
\AUTHOR{Feiyang (Amber) Xu, Poonacha K. Medappa, Murat M. Tunc \\ Martijn Vroegindeweij, Jan C. Fransoo}
\AFF{Tilburg University, the Netherlands \\ \EMAIL{f.xu\_1@tilburguniversity.edu, p.k.medappa@tilburguniversity.edu, m.m.tunc@tilburguniversity.edu \\ w.m.vroegindeweij@tilburguniversity.edu, jan.fransoo@tilburguniversity.edu}}
}

\ABSTRACT{

GenAI solutions like GitHub Copilot have been shown to increase the productivity of software developers. Yet prior work remains unclear on the quality of code produced and the challenges of maintaining it in software projects. If quality declines as volume grows, technical debt accumulates as experienced developers face increased workloads reviewing and reworking code from less-experienced contributors. We analyze developer activity in Open Source Software (OSS) projects following the introduction of GitHub Copilot. We find that productivity indeed increases. However, the increase in productivity is primarily driven by less-experienced (peripheral) developers. We also find that code written after the adoption of AI requires more rework to satisfy repository standards, indicating a potential increase in technical debt. Importantly, the added rework burden falls on the more experienced (core) developers, who review 6.5\% more code after Copilot’s introduction, but show a 19\% drop in their original code productivity. More broadly, this finding raises caution that productivity gains of AI may mask the growing burden of maintenance on a shrinking pool of experts, together with increased technical debt for the projects. The results highlight a fundamental tension in AI-assisted software development between short-term productivity gains and long-term system sustainability.

}

\maketitle
\vspace{-1cm}

\section{Introduction}
\hfill

How will AI shape the future of knowledge-intensive industries? This question has taken on renewed significance with the recent rise of Genarative AI (GenAI) technologies, which are becoming an integral part of daily operations of software development, scientific research, healthcare and other expert-driven fields. A prominent example is GitHub Copilot, an AI-powered coding assistant designed to support developers by generating code suggestions and accelerating routine programming tasks \citep{Peng2023}. When GitHub launched Copilot, it was introduced as “your AI pair programmer," emphasizing not only its role as an automation tool but also as a team member who partners with the developer to create knowledge \citep{Friedman2021}. 
Unlike earlier coding automation tools that primarily targeted productivity, GitHub Copilot’s framing as a pair programmer signals a deeper shift. It implies that AI may fundamentally reshape how knowledge-intensive work is performed, coordinated, and organized, rather than merely accelerating existing tasks.

For organizations and communities involved in software development, the addition of AI pair programmers in teams offers the potential for significant productivity gains. Right after the launch of GitHub Copilot, research shows that developers who use Copilot completed their programming tasks 55.8\% faster \citep{Peng2023}. Such productivity benefits lead to promises of faster time-to-market and increased revenue for organizations developing software applications. 
Considering these shifts, major tech organizations have started to increasingly rely on AI in their projects - “more than a quarter of all new code at Google is generated by AI, then reviewed and accepted by engineers," reported Google CEO Sundar Pichai in January, 2025.\footnote{https://www.technologyreview.com/2025/01/20/1110180/the-second-wave-of-ai-coding-is-here/} Moreover, Microsoft CTO Kevin Scott expects that 95\% of all code will be AI-generated by 2030.\footnote{https://www.businessinsider.com/microsoft-cto-ai-generated-code-software-developer-job-change-2025-4}

While these productivity gains are promising, they also raise important questions about the quality and maintainability of AI-generated code. Because AI tools can lower the skill barrier for writing code \citep{dakhel2023}, AI tools enable broader participation but may also encourage developers to rely on generated solutions without fully understanding the underlying design rationale and potential integration issues \citep{Barrett2023}. 
Such reliance increases the likelihood of quick fixes that favor short-term functionality over long-term maintainability \citep{Barrett2023}.
Extant literature characterizes “quick and dirty” software customizations made without a complete understanding of their future implications as technical debt, as they undermine system reliability and impose long-term maintenance obligations \citep{kructenIEEE,brownACM,Banker2021}. 
As a result, project maintainers must devote additional effort to understanding, reviewing, and reworking AI-generated code before it can be safely integrated. In our context, we contend that the growing reliance on AI-assisted development may accelerate the accumulation of technical debt, as design shortcuts taken to expedite system deployment become embedded in software systems \citep{Ramasubbu2016, RamasubbuKemerer2021}.

The technical debt and maintenance challenges that AI poses are expected to be especially pronounced in distributed software development teams, such as in Open Source Software (OSS) communities. In these communities, contributors from around the world collaborate, often voluntarily, to develop and maintain software that form the digital infrastructure of our society (e.g., Linux, Apache, LaTeX, Python), making it freely or cheaply available to the public \citep{eghbal2020, Nagle2019MS}.  Despite the voluntary nature of work in these communities, OSS constitutes critical digital infrastructure for modern society, with estimates suggesting that the total cost of reproducing this software would amount to \$8.8 trillion \citep{Hoffmann2024}.\footnote{These estimates suggests that firms would spend approximately 3.5 times more on software than they currently do if OSS did not exist \citep{Hoffmann2024}.} Given this critical role for both firms and society, the growing adoption of AI in OSS communities raises important questions about its broader impact. On the one hand, AI tools can lower the barrier for peripheral contributors \citep[relatively newer contributors to the project who come from the community of users of software; ][]{rullani2012} to contribute to these software projects. On the other hand, this surge in AI-assisted contributions may result in a build up of technical debt and increase the maintenance obligations of the software \citep{Banker2021}. If this mechanism holds, the performance consequences of technical debt are likely to intensify over time, as mounting maintenance burdens increasingly divert expert attention away from productive development activities, thereby amplifying the negative impact of technical debt on project and firm performance. 

This study seeks to address this tension. We can distinguish two important types of activities in OSS projects - development and maintenance. The development activity in OSS projects involves producing code and submitting pull requests (PRs) to the project. The open nature of OSS communities allows anyone—regardless of skill or experience - to contribute to this process \citep{rullani2012}. In contrast, maintenance involves reactive tasks such as triaging issues, reviewing submitted PRs, suggesting modifications or corrections, and ensuring that contributions align with project standards \citep{Banker2021, RamasubbuKemerer2021}. These maintenance tasks are generally less intrinsically driven and typically assigned to contributors who possess both technical expertise and trust of the community to ensure the quality and reliability of the codebase \citep{medappa2023sponsorship, eghbal2020}. Thus, we seek to answer the question: \textit{How does AI-assisted programming impact the development and maintenance activities and the technical debt of the OSS projects?}

In this study, we examine whether technical debt and maintenance efforts of OSS projects changed after the introduction of GitHub Copilot through increased code review and rework effort on PRs. To empirically test this, we exploit the release of GitHub Copilot as a technical preview in June 2021, which included limited programming language endorsement. We focus on OSS projects owned by Microsoft, as the company had exclusive access to OpenAI's GPT-3, the model powering GitHub Copilot during its technical preview, due to its investment in OpenAI and its prior acquisition of GitHub.\footnote{https://www.technologyreview.com/2020/09/23/1008729/openai-is-giving-microsoft-exclusive-access-to-its-gpt-3-language-model/}\footnote{https://www.mckinsey.com/industries/technology-media-and-telecommunications/our-insights/thomas-dohmke-on-improving-engineering-experience-using-generative-ai} The individual users in our dataset are contributors to Microsoft-owned OSS projects. We estimate the effect of Copilot at both the project and contributor levels using a Difference-in-Differences (DiD) design. Treatment and control groups were defined based on the primary programming language: those using Copilot-endorsed languages formed the treatment group, while non-endorsed language users served as the control \citep{yeverechyahu2024impact}. For both project and contributor levels, we collected data on programming activities and aggregated them at the monthly level.

We examine the changes in code productivity after Copilot by three measures: lines of code added, commits \footnote{A commit is the fundamental unit of change on GitHub. Similar to saving a file that's been edited, a commit records changes to one or more files on GitHub - https://docs.github.com/en/pull-requests/committing-changes-to-your-project/creating-and-editing-commits/about-commits} and PRs submitted to the project. To capture the secondary effects of AI adoption on technical debt and maintenance effortsof OSS communities, we focused on two complementary outcomes. First, we measured technical debt at the project level using PR rework, which reflects the extent to which initially submitted contributions require modification before being integrated into the codebase \citep{Ramasubbu2016, RamasubbuKemerer2021}. Higher levels of rework indicate greater reliance on expedient or insufficiently integrated solutions. Second, we measured maintenance effort at the individual level using PR reviews, which capture the time and effort required from contributors to evaluate, correct, and integrate submitted code \citep{medappa2023sponsorship}. 

Based on our analysis of a large-scale panel dataset from GitHub, we find that while AI adoption leads to productivity gains, they also increase maintenance-related activities due to a higher volume of review and rework needed per PR. Specifically, our analysis reveals a double-edged effect of GitHub Copilot on OSS development. At the project level, Copilot adoption is associated with a significant boost in productivity: projects that supported Copilot saw increases in lines of code added, commits, and PRs. However, this surge in contributions came also with an increase in PR rework (2.4\% more code revisions), indicating a possible decline in the quality of code initially submitted resulting in an accumulation of technical debt. At the contributor level, we observe an important redistribution of effort: the peripheral contributors (the less active contributors to the projects) increased their development activity, taking advantage of Copilot’s ability to lower coding barriers. Specifically, peripheral contributors, particularly those in the bottom percentiles in terms of their previous contributions, increased their commit activity by 43.5\% and submitted 17.7\% more PRs. In contrast, the core contributors reduced their commit activity by 19\%, shifting their focus toward reviewing and maintaining code (a 6.5\% increase), and shouldering a heavier quality assurance burden. Together, these findings highlight how AI can enable broader participation in OSS, but also raise concerns about the sustainability of these gains and the strain placed on a shrinking pool of experienced contributors who maintain quality in OSS projects.

\section{Related Literature and Conceptual Development}
\hfill

\subsection{AI Assisted Code Development in OSS Projects}
\hfill

The rapid advancement of GenAI technologies, particularly LLMs like ChatGPT and GitHub Copilot, is transforming how software is developed and how online knowledge communities operate. Current research on AI-assisted code development has shown the substantial impact of the technology on productivity. A study to assess the productivity benefits of using Copilot revealed that developers who use Copilot completed their programming task 55.8\% faster than the control group \citep{Peng2023}. Another study by GitHub reports that the use of Copilot Chat increases programmers’ confidence, with participants self-reporting improvements in code readability, reusability, conciseness, maintainability, and resilience \citep{Github2023}. These productivity gains also translate into labor-market outcomes: developers exposed to AI-assisted coding experience faster career progression in the short- to medium-term \citep{li2025exploring}. Work also finds that AI coding assistants reshape the allocation of work. For instance, \cite{yeverechyahu2024impact} investigate the impact of GitHub Copilot on innovation in OSS projects. They find a significant increase in overall code contributions, accompanied by a shift in the nature of innovation toward more routine and incremental changes. \cite{song2025impactgenerativeaicollaborative} find that Copilot adoption increases project-level code contributions, though this comes at the cost of an increase in coordination time for code integration. Relatedly, \cite{hoffmann2025generative} show that access to GitHub Copilot reallocates developers' effort toward core coding tasks and away from project management and coordination activities.

While AI-assisted code development promises substantial productivity gains, its implications for software maintenance remain less well understood. Prior research on software development has long recognized that development costs are often small relative to maintenance costs, which include sustaining activities associated with ensuring software quality and security \citep{Nagle2019MS}.  In the case of OSS, while users can benefit from reduced up-front costs, collective intelligence of the crowd, and flexibility to implement changes, the challenges of maintenance get magnified as contributors are not contractually obligated to maintain the software \citep{VonHippel2003OrgSc, Nagle2019MS}. The Linux Foundation's OSS Contributor Survey provides insightful perspectives on the complexities involved in maintaining OSS \citep{linuxsurvey}. Firstly, it highlights that ``general housekeeping" tasks, such as project maintenance, bug reporting and documentation, and organizational or administrative duties, often consume a more significant portion of contributors' time than desired. Secondly, despite a preference among contributors to spend less time on maintenance tasks, there's a broad acknowledgment of the importance of these activities, especially those related to software security, for the success and integrity of their projects \citep{linuxsurvey}.

Furthermore, AI code assistants, including prompt-based and “vibe coding” practices, promise to increase productivity while easing access for contributors to submit code, even in complex and mature OSS projects. Recent work has begun to examine vibe coding as an emerging and controversial paradigm in AI-assisted software development, in which programmers rely on natural language interaction with generative models to maintain flow and rapidly explore solutions, often with minimal upfront specification \citep{Pimenova2025, FawzyTahirBlincoe2025}. While this approach can substantially accelerate development and foster experimentation, the literature consistently highlights associated risks, including underspecified requirements, reduced reliability, difficulties in debugging, increased latency, and heavier burdens on code review and collaboration \citep{HeMillerAgarwalKastnerVasilescu2025}. A recurring theme is a speed–quality paradox \citep{FawzyTahirBlincoe2025}: although vibe coding enables rapid progress and is often perceived as highly efficient by its users, the resulting code is frequently described as fast but flawed. Moreover, quality assurance practices are commonly deprioritized in vibe coding workflows, with users skipping testing, accepting AI-generated outputs with little modification, or delegating verification back to the AI itself. These patterns raise concerns about latent defects and heightened maintenance demands, particularly in collaborative and production-oriented software environments \citep{Schreiber_2025}. 

These maintenance-related concerns are amplified by evidence that less experienced programmers appear to benefit more from the use of Copilot. For novice developers, AI-driven code suggestions can help them overcome steep learning curves and contribute even to mature OSS projects. However, both \cite{dakhel2023} and \cite{Peng2023} caution that novice programmers may place undue trust in AI-generated code without fully understanding its broader design or integration implications, increasing the likelihood of downstream quality and integration challenges.
Over time, these issues can accumulate as technical debt, increasing the maintenance burden for OSS projects and shifting responsibility toward a small group of experienced core contributors. 
Anecdotal evidence\footnote{https://stackoverflow.blog/2025/08/07/a-new-worst-coder-has-entered-the-chat-vibe-coding-without-code-knowledge/} suggests that while AI-assisted coding can produce seemingly complete solutions with minimal effort, these outputs frequently exhibit substantive deficiencies when reviewed by experienced developers. Furthermore, \cite{Schreiber_2025} show that even when AI-generated code passes functional tests, it often embeds latent security weaknesses and maintainability issues that shift effort from development to downstream debugging and remediation, effectively accumulating technical debt. Relatedly, the growing volume of low-quality, AI-generated security artifacts has begun to strain OSS security workflows. In January 2026, the cURL open-source project discontinued its bug bounty program after being overwhelmed by a flood of AI-generated vulnerability reports that consumed reviewer attention without yielding actionable findings \footnote{https://www.techradar.com/pro/security/curl-will-stop-bug-bounties-program-due-to-avalanche-of-ai-slop}. In other words, the ease of generating code masks underlying structural and logical gaps that become apparent under expert scrutiny \citep{Barrett2023}. As with other AI tools that create superficially polished but potentially shallow results, vibe coding may be appealing for rapid prototyping, yet the secondary effects that this may have on reliability or maintainability when applied to nontrivial software development tasks is an open question.

Our study investigates this issue by analyzing how the adoption of GitHub Copilot reshapes contributor effort across production and maintenance activities, and whether it disproportionately shifts the burden of maintenance onto a smaller group of core contributors.

\subsection{Conceptualization of Technical Debt}

Technical debt captures the intertemporal trade-off between short-term development speed and long-term system maintainability \citep{Cunningham1992}. A growing body of research has established that technical debt is economically consequential, persistent, and shaped by organizational choices. Drawing on the information systems and computer science literature, we synthesize prior work in Table \ref{tab:techdebt_lit} and adopt a definition of technical debt in this study as accumulation of suboptimal code, design, or documentation decisions in PR process that increase PR Rework (technical debt) and PR Review (maintenance effort) - a definition that is particularly prominent in the era of GenAI software development. 

Measurement approaches for technical debt have evolved along two complementary axes. One strand emphasizes static code metrics (e.g., code complexity, duplication, test coverage) as proxies for debt accumulation, enabling automated detection and cross-project comparisons \citep{ParamithaMassacci2023, YooCraigheadSamtani2025}. \cite{ParamithaMassacci2023} show how dependency structures in open-source ecosystems (top 600 Python packages) amplify technical leverage, allowing localized weaknesses to propagate across packages. \cite{YooCraigheadSamtani2025} similarly demonstrate that dependency network structures shape the diffusion of security vulnerabilities in software supply chains. Another strand operationalizes technical debt through process and outcome indicators—such as rework rates, defect density, time-to-merge, and maintenance effort—thereby connecting debt to developer behavior and organizational performance \citep{Ramasubbu2016, Banker2021, RamasubbuKemerer2021}. Both approaches have value: the former metric-based techniques are scalable and actionable for tooling, while the later process indicators capture the economic and human costs that managers and maintainers ultimately face.

Empirical studies across industry and OSS ecosystems consistently find that technical debt impairs long-run productivity and increases maintenance burdens \citep{ParamithaMassacci2023}. Work in empirical software engineering shows that higher measured debt correlates with increased bug rates, longer defect resolution times, and reduced velocity for feature delivery \citep{Ramasubbu2016}. \cite{RintaKahila2023} further demonstrate that organizations can become “trapped” in technical debt due to coordination failures, organizational inertia, and escalating switching costs. The introduction of GenAI tools such as GitHub Copilot represents a departure from these assumptions \citep{yeverechyahu2024impact}. Unlike prior productivity-enhancing tools, GenAI dramatically lowers the marginal cost of producing code and reduces skill barriers to contribution \citep{Peng2023}. While prior studies document productivity gains from AI-assisted coding, they provide limited insight into how these gains translate into technical debt and maintenance related challenges. 

From a technical debt perspective, GenAI may accelerate debt accumulation by increasing code volume without proportionate improvements in quality, the integration with the software, or architectural coherence \citep{Pimenova2025}. By lowering the cost of producing code, AI-assisted programming encourages rapid iteration and experimentation, but can shift attention away from longer-term concerns such as maintainability, readability, and alignment with existing system design \citep{Barrett2023, Schreiber_2025}. As a result, defects, inconsistencies, and design shortcuts may be introduced more quickly than they can be identified and resolved. Over time, this imbalance can compound, transforming short-term productivity gains into persistent maintenance obligations that must be absorbed by experienced developers \citep{eghbal2020}. While GenAI promises to enhance development speed and broaden participation, it may simultaneously intensify the very technical debt that constrains system reliability, scalability, and long-run performance \citep{RamasubbuKemerer2021}. If this logic is valid, we expect that GenAI may lead to a greater accumulation of technical debt in OSS projects.

\newpage
\begin{landscape}

\begin{table}[h!]
\centering
\caption{Selected Literature on Technical Debt and Software Maintenance}
\label{tab:techdebt_lit}

\resizebox{\linewidth}{!}{%
\begin{tabular}{p{2cm} p{4cm} p{3.8cm} p{6.6cm} p{6cm}}
\toprule
\textbf{Study} & \textbf{Context} & \textbf{Method} & \textbf{Measurement} & \textbf{Key Findings} \\
\midrule
Banker et al. (2021) & Customer relationship management (CRM) systems in 26 firms & Econometric analysis & Percentage of customized codes in the CRM system that do not adhere to vendor-prescribed standard & Higher technical debt is associated with lower firm performance and reduced operational efficiency over time. \\
Ramasubbu \& Kemerer (2021) & Outsourced Commercial Off-The-Shelf (COT) enterprise systems & Econometric analysis & Violations of the design and programming standards established by the vendor of the COTS enterprise system & Active remediation policies reduce long-term maintenance costs, but excessive deferral leads to escalating technical debt. \\
Ramasubbu \& Kemerer (2016) & Enterprise software systems & Econometric analysis & Client customizations 
violating vendor standards (business logic / data schema), API violations & Accumulated technical debt increases system failure hazards and shortens system lifespan. \\
Paramitha \& Massacci (2023) & Top 600 Python packages, Python open-source ecosystem & Box plot, simulation & Technical leverage as the ratio between dependencies (other people's code) and own codes of a software package & Technical leverage amplifies maintenance risks through dependency networks, increasing downstream technical debt. \\
Yoo et al. (2025) & Software supply chains & Econometric analysis & Dependency centrality, vulnerability exposure & Complex dependency structures increase vulnerability propagation, linking technical debt to security risks. \\
Rinta-Kahila et al. (2023) & Legacy system replacement & Qualitative case study, system-dynamics model & Sociotechnical debt indicators: architectural compromises, technical inertia \& digital options foregone or constrained & Technical debt becomes entrenched through organizational routines, creating lock-in and delaying modernization. \\
Rolland et al. (2018) & Digital platforms in firms & Longitudinal case analysis & The interactions between digital options and digital debt as a contribution to understanding the complex choices organization face in managing digital platforms & Strategic flexibility often trades off with accumulating digital and technical debt. \\
This study & OSS Repositories on GitHub & Econometric
analysis & Accumulation of suboptimal code, design, or documentation decisions in PR process that increase PR Rework (technical debt) and PR Review (maintenance effort) & Code written after the adoption of AI requires more rework to satisfy repository standards \\
\bottomrule
\bottomrule
\end{tabular}
}
\end{table}

\end{landscape}

\subsection{Technical Debt in OSS projects}
Prior research on software quality and technical debt has largely focused on proprietary software developed within organizational boundaries, where development and maintenance are often treated as sequential and centrally coordinated activities \citep{ParamithaMassacci2023, YooCraigheadSamtani2025, Ramasubbu2016, Banker2021, RamasubbuKemerer2021}. In contrast, OSS is produced through a distributed development model in which participation is open and contributions can be submitted by anyone, blurring the boundary between development and maintenance and creating distinct challenges for managing quality and technical debt \citep{Nagle2019MS}. Therefore, our study focuses on the PR review process, a core coordination mechanism in continuous integration and continuous deployment (CI/CD) of software. 

\begin{figure}[htbp]
\centering
\includegraphics[width=0.6\textwidth]{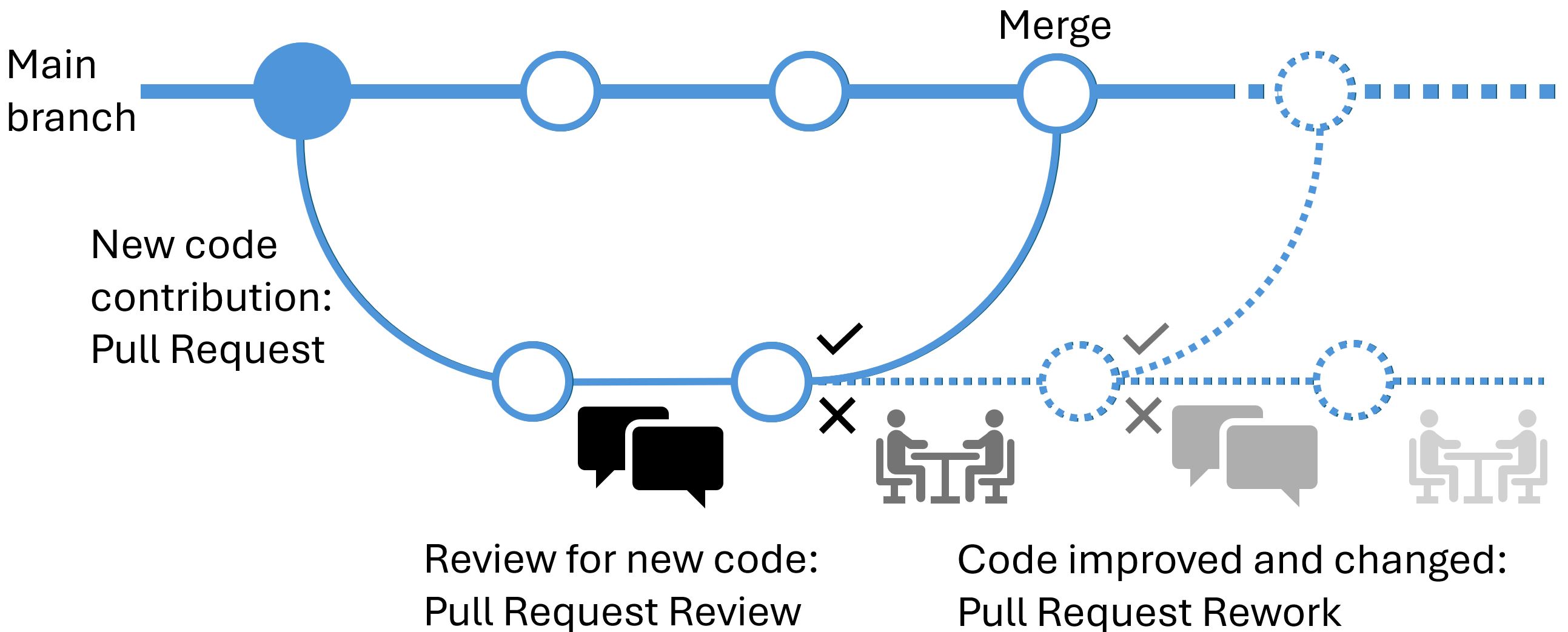}
    \caption{The workflow of OSS projects. It comprises of a primary branch of a GitHub  (project), which typically contains the main source code that serves as the foundation for new feature development, bug fixes, and updates. Changes to this branch are usually controlled through a structured review process conducted by maintainers to ensure code quality and prevent issues.To contribute to a project, contributors submit a PR to merge their code changes, such as new features or improvements. It allows developers to submit modifications, request feedback, and merge updates into the main branch. The review and revision process comprises of two activities that we measure in our study -  PR review and PR rework. Eventually the reviewed and reworked changes will either be merged into the main branch or undergo additional rounds of review and revision.}
    \label{fig:PR_diagram}
\end{figure}

The development workflow of an OSS project is illustrated in Figure \ref{fig:PR_diagram}. A feature of this development workflow is that it allows multiple contributors to work independently on separate branches (or forks) and submit their changes for inclusion in the main project branch. This is done through a PR, which serves as a formal request to review and merge code contributions (such as new features or improvements) made by project contributors. Core or experienced contributors often handle key maintenance tasks: they review submitted PRs for quality and compliance, suggest improvements or corrections adhere to project standards \citep{Banker2021}, and integrate approved changes into the main project. This workflow enables distributed development and community-driven innovation, allowing peripheral contributors who come from the community of users of the project to contribute. At the same time, it ensures that the code is reviewed, refined, and improved before merging into the main branch. The effectiveness of this review process has helped OSS achieve remarkably high quality, surpassing proprietary software in metrics like bugs per 1,000 lines of code.\footnote{\url{https://blogs.worldbank.org/en/opendata/quality-open-source-software-how-many-eyes-are-enough}} However, the success and effectiveness of this "decentralized-development" approach hinges on the core contributors who actively participate in the review-rework process of the PRs submitted by different contributors \citep{rullani2012}. 

Focusing on the PR review process is particularly well suited to OSS contexts, where code is continuously developed, reviewed, and deployed into production, often with rapid and frequent version updates \citep{Roberts2006MS, medappa2023sponsorship}. By analyzing PR reviews and rework as they occur, our study captures maintenance activities in real time, providing a more accurate representation of how quality control and technical debt are managed within distributed development workflows. Importantly, the PR review process allows us to observe technical debt at the point of entry—when new code is evaluated, revised, or rejected prior to integration. This technical debt measurement would be applied to other CI/CD software projects beyond OSS community.

Prior work conceptualizes technical debt as deferred maintenance or suboptimal design choices that increase future rework costs, typically operationalized through downstream quality degradation or maintenance outcomes in traditional, organization-bound software development settings \citep{Ramasubbu2016, Banker2021, RamasubbuKemerer2021}. In contrast, OSS development unfolds through continuous integration and deployment of new features, where design trade-offs and quality concerns are often raised and resolved at the PR level. In this context, PR rework provides a direct operationalization of technical debt, capturing realized remediation effort during code integration.


\section{Data and Methods}
\hfill
\subsection{Research Design}

\begin{figure}[H]
\centering
\begin{tikzpicture}[scale=1.2]
\draw [thick,->] (-6,0) -- (3,0);
\draw (-6, 0) node[font=\footnotesize\linespread{1}\selectfont,align=center,above=3pt] {Data starts} ;
\draw (-6, 0) node[font=\footnotesize\linespread{1}\selectfont,align=center,below=3pt] {July 2020} ;
\draw (-2, 0) node[font=\footnotesize\linespread{1}\selectfont,align=center,above=3pt] {GitHub Copilot  \\ Technical Preview };
\draw (-2, 0) node[font=\footnotesize\linespread{1}\selectfont,align=center,below=3pt] { June 29, 2021};
\draw (1.3, 0) node[font=\footnotesize\linespread{1}\selectfont,align=center,above=3pt] { Data ends };
\draw (1.3, 0) node[font=\footnotesize\linespread{1}\selectfont,align=center,below=3pt] { July 2022};
\draw [font=\footnotesize\linespread{1}\selectfont,decorate,decoration={brace}] (-6,1) -- (-2.05,1) node[midway,yshift=1em]{Pre-Treatment Period};
\draw [font=\footnotesize\linespread{1}\selectfont,decorate,decoration={brace}] (-1.95,1) -- (1.3,1) node[midway,yshift=1em]{Post-Treatment Period};
\draw (-1.5, -0.6) node[font=\footnotesize\linespread{1}\selectfont,align=center,below=3pt] {\textbf{Treatment Group}: Python, JavaScript, Ruby, TypeScript and Go};
\draw (-2, -1) node[font=\footnotesize\linespread{1}\selectfont,align=center,below=3pt] {\textbf{Control Group}: R, C, C\#, C++, Java, PHP and Scala};
\draw[thick] (-6,-0.1) -- (-6,0.1);
\draw[thick] (-2,-0.1) -- (-2,0.1);
\draw[thick] (1.3,-0.1) -- (1.3,0.1);
\end{tikzpicture}
\caption{The timeline of our study period
}
    \label{fig:releasedatesAI}
\end{figure}

In June 29, 2021, developed on OpenAI’s GPT-3 model, GitHub launched a technical preview version of Copilot\footnote{In September 2020, Microsoft gained exclusive access to OpenAI's GPT3. This access allowed Microsoft to repurpose and modify the model for code generation, leading to the development of GitHub Copilot. In June 2021, GitHub Copilot was launched as a technical preview. Anecdotal evidence and our interviews of Microsoft employees indicates that during the technical preview, access to Copilot was restricted to selected GitHub users, specifically employees of Microsoft/GitHub and maintainers of popular projects. This restricted access suggests that the primary users of Copilot during the technical preview were likely Microsoft and GitHub employees, along with selected Github uses who volunteered for the beta testing the software.}, an AI-powered LLM designed to assist with coding. This early version of GitHub Copilot was not open to the public and endorsed five programming languages: Python, JavaScript, Ruby, TypeScript, and Go.\footnote{https://github.blog/news-insights/product-news/introducing-github-Copilot-ai-pair-programmer/} Copilot was later launched to the public in June 2022, with contributors required to pay a monthly fee to subscribe to the AI pair programming service. With the public release in June 2022, Copilot gradually added endorsement for more languages such as, C and Java. As shown in Figure \ref{fig:releasedatesAI}, we define our observation period as 12 months before and 12 months after its introduction as technical preview. The measures were aggregated to create a monthly panel dataset spanning from July 2020 to July 2022. 

Our study leverages the natural experiment created by the launch of GitHub Copilot as a technical preview. Specifically, we exploit Copilot’s early-stage language endorsement, which included Python, JavaScript, Ruby, TypeScript, and Go, while excluding other comparable languages such as R, C, C\#, C++, Java, PHP, and Scala. We select these non-Copilot-endorsed languages as the control group because they offer comparable functionality and are among the most frequently used languages in Microsoft-owned repositories. Our identification strategy, which contrasts programming languages across treatment and control groups, is consistent with prior studies \citep{yeverechyahu2024impact}.

Anecdotal evidence\footnote{https://www.mckinsey.com/industries/technology-media-and-telecommunications/our-insights/thomas-dohmke-on-improving-engineering-experience-using-generative-ai}  and our interviews with Microsoft employees indicates that during the technical preview, access to Copilot was restricted to selected GitHub contributors, specifically employees of Microsoft/GitHub and maintainers of popular repositories who volunteered to participate in the restricted preview of the tool. This restricted access suggests that the primary users of Copilot during the technical preview were likely Microsoft and GitHub (organization) employees, ensuring that the tool was tested internally before its broader release.\footnote{Dog Fooding; is Microsoft speak for internal use of their own software to ensure it is tested before public release https://devblogs.microsoft.com/oldnewthing/20110802-00/?p=10003}

Since Copilot usage cannot be identified at the individual level, we focus on Microsoft-owned repositories and contributors who actively contributed to these repositories during the observation period. By doing so, we aim to capture changes in OSS contributor behavior driven by Copilot, as contributors to Microsoft-owned repositories are more likely to have access to the tool during the technical preview period.

\subsection{Data Collection}

We use GitHub's API service to collect all the data for this research, enabling efficient and precise querying of repository / individual activities. This approach enables us to gather detailed information on measurements such as PRs, commits, reviews, authors, and repository / individual metadata, providing a comprehensive dataset for analyzing the impact of AI-assisted coding on OSS project development and maintenance activities.
 
At the project level, we define the treatment group as repositories whose primary programming language was among those endorsed by GitHub Copilot during the technical preview, while repositories using non-endorsed languages (e.g., C, C\#, C++, Java, PHP, R, Scala) serve as the control group. Our project-level dataset consists of 2,755 repositories, with 1,660 in the Copilot-endorsed treatment group and 1,095 in the non-endorsed control group.

Our individual level dataset consisting of 1,699 contributors from GitHub, with 1,186 in the Copilot-supported treatment group and 513 in the non-Copilot-supported control group.\footnote{There are 37,334 contributors who made at least one contribution to the repositories in our sample. The contributions include, for example, posting a comment, submitting a commit, or conducting a PR review. Among them, we filtered out 5,308 contributors who participated in more than three of the repositories we studied (we selected three repos to ensure sufficient variation for our PR reviewed repositories measure) . Then, we applied a programming language filter to construct a comparable treatment and control group for the individual-level analysis, and the data set was reduced to 1,699 contributors (who were users of the treated and control programming languages).}

\subsection{Variables}
Table \ref{tab:var_repo} presents the main variables and provide descriptive statistics. 
Our dataset is aggregated to the repository-month / contributor-month level, summary statistics are calculated based on repository-month / contributor-month observations.

\begin{table}[h!]
\centering
\caption{Variable Descriptions}
\label{tab:var_repo}
\begin{tabular}{p{5cm} p{10cm}}
\hline
\textbf{Variable} & \textbf{Description} \\
\hline

\multicolumn{2}{l}{\textbf{Dependent Variables}} \\[2pt]
Code Added (log) & 
Line of codes submitted to a repository in a given month. \\
Commits (log) & 
Commits submitted to a repository (by a contributor) in a given month. On GitHub, each commit typically represents a completed piece of work, such as fixing a bug, adding a feature, or improving existing code.\\
PRs (log) & 
Pull requests submitted to a repository (by a contributor) in a given month. Each pull request represents a bundle of proposed changes (commits) that must be reviewed and approved before becoming part of the repository.\\

Techinal Debt & 
Measured by PR rework (log), count of follow-up commits added to a PR after initial submission, capturing the extent of required code revision. \\

Maintenance Efforts& 
Measured by PR review (log), count of PR reviews conducted by a contributor in a given month, analyzed at individual level.  \\[6pt]

\multicolumn{2}{l}{\textbf{Independent Variables}} \\[2pt]

Treatment & 
Indicator variable equal to 1 if the repository’s primary programming language is one of the five Copilot-supported languages, and 0 otherwise. \\

Copilot & 
Indicator variable equal to 1 for months after the introduction of GitHub Copilot (July 2021), and 0 otherwise. \\

Treatment $\times$ Copilot & 
Interaction term capturing the DiD estimate of Copilot’s effect on repository-level development and maintenance outcomes. \\[6pt]

\multicolumn{2}{l}{\textbf{Control Variables}} \\[2pt]

PRs (log) & 
Included as a control in PR rework regressions to account for heterogeneity in code submission volume across repositories. \\

Month-year fixed effects & 
Indicator variables capturing temporal shocks common across all repositories (e.g., seasonal patterns). \\

Repository fixed effects & 
Repository-level constants controlling for time-invariant heterogeneity (e.g., project age, governance structure). \\

\hline
\end{tabular}
\begin{tablenotes}
\emph{Note:} All DVs are log-transformed. A value of one is added prior to transformation to address zeros.   
\end{tablenotes}
\end{table}

\subsubsection{Dependent Variable: PR Rework}
\hfill

In this study, we operationalize technical debt through process and outcome indicators, using rework rates, and maintenance effort, thereby connecting debt to developer behavior and organizational performance \citep{Banker2021, RamasubbuKemerer2021}. Specifically, \cite{Banker2021, RamasubbuKemerer2021} measure technical debt as the proportion of customized code in customer relationship management (CRM) systems that does not adhere to vendor-prescribed standards. Analogously, in the OSS context, we measure technical debt by the number of commit amendments (PR rework) required for a PR to satisfy repository standards, capturing the additional rework effort induced by nonconforming code.

On GitHub, a PR is a vehicle of contribution through which contributors participate in the development process. A PR typically contains one or more commits and often represents a "patch" or feature addition to the project. Any individual can submit a PR (a request to merge their contribution into the project), which is then (peer) reviewed by the core contributors, who have write access to the source code. The core contributors reviewing the PR can decide to merge the PR into the main project, request modifications or, reject it. If modifications are requested for a PR, the author of the PR can address the comments and modification requests and re-submit the code in the form of follow-up commits for another round of review. This revision process continues until all issues with the code are resolved and the code can be merged, or until the idea behind the code is no longer in alignment with the project goals and the PR is rejected. Taking this into account, we can measure the extent of rework done on a PR submitted by a contributor by identifying the number of commits that are added to the PR after its initial submission. We use the PR rework measure as a proxy to determine the technical debt attributable to initial code contributions submitted to the project and the maintenance-related efforts associated with the code contributions.

\subsubsection{Dependent Variable: PR Review and PR Reviewed Repos}
\hfill

As described in the previous section, we use PR rework as a proxy for the accumulation of technical debt. Consistent with prior literature \citep{Banker2021, RamasubbuKemerer2021}, we further conceptualize PR reviews and PR reviewed repositories as measures of maintenance effort aimed at preventing or mitigating technical debt.

PR reviews capture the intensity of review activity undertaken by a contributor and are defined as the number of PR for which the contributor provides formal review during the observation period. This measure reflects the effort devoted to evaluating, debugging, and improving submitted code prior to integration.

PR reviewed repositories capture the breadth of a contributor’s maintenance responsibilities and are defined as the number of distinct repositories in which the contributor conducts at least one PR review during the observation period. This variable reflects the extent to which maintenance effort is distributed across multiple projects and repositories.

\subsubsection{Moderation Variable: Core Contributors}
\hfill

\begin{figure}[H]
    \centering
    \begin{subfigure}{0.3\textwidth}
        \centering
        \includegraphics[width=\textwidth]{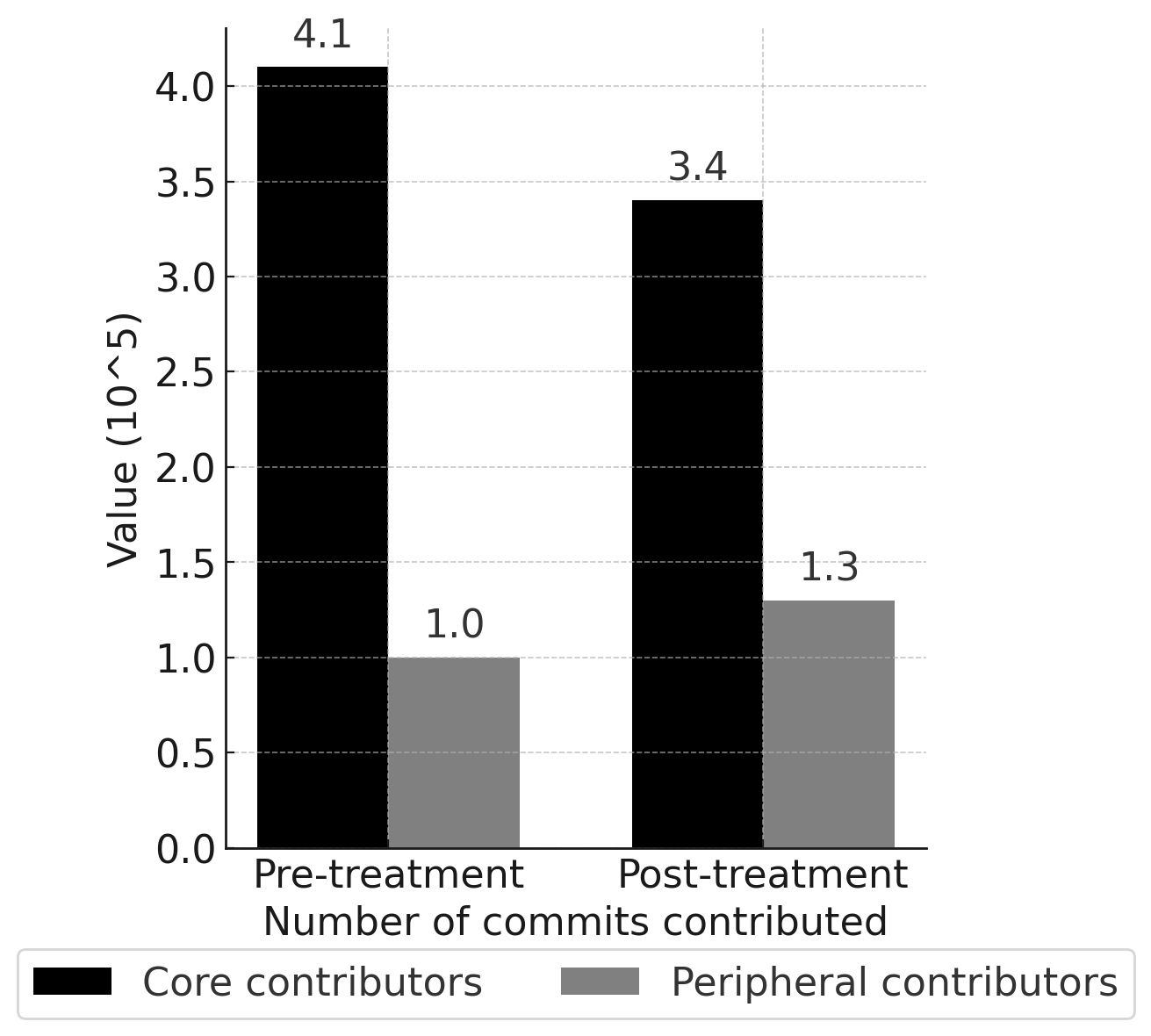}
    \end{subfigure}
    \begin{subfigure}{0.3\textwidth}
        \centering
        \includegraphics[width=\textwidth]{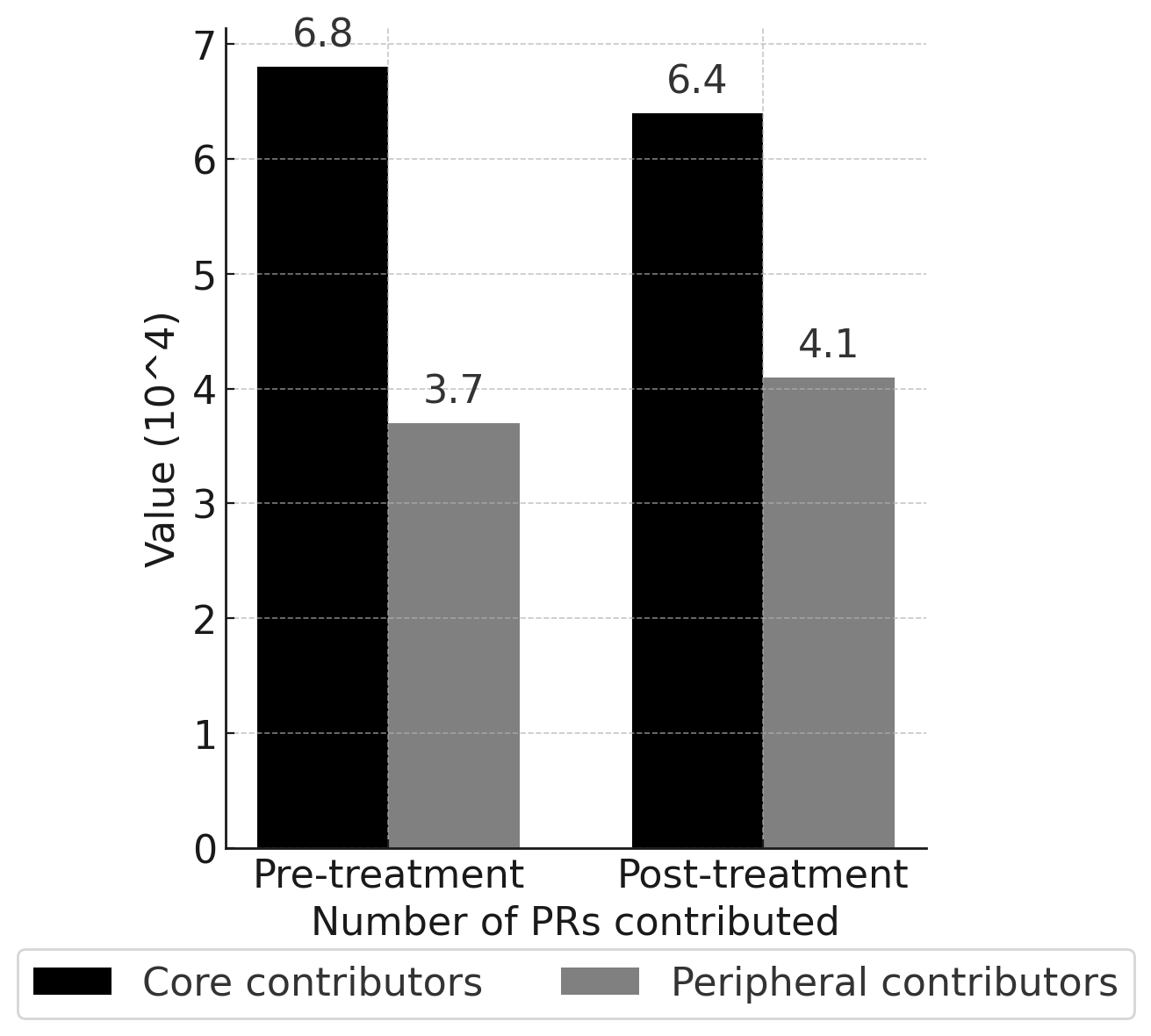}
    \end{subfigure}

    \caption{Histogram of contributions for core and peripheral contributors: The core contributors decreased the development activities after the deployment of Copilot. The peripheral contributors displayed opposite behaviour.}
    \label{fig:3x1graphs}
\end{figure}

For the conceptualization of core contributors we used contributors' level of activity during the pretreatment period, measured by the number of commits, to classify contributors into core contributors (top 25\%) and peripheral contributors (rest 75\%) \citep{Setia2012, crowston2006core}. The classification of contributors allows us to examine whether the effect of Copilot varies depending on a core versus a periphery contributor. 
From Figure \ref{fig:3x1graphs}, we observe that core contributors perform the majority of development activities (both in terms of commits and PR) in the projects. 

\subsubsection{Descriptive Statistics}
\hfill

We report summary statistics for the key variables used in our analyses. Table \ref{tab:summary_stats} presents the descriptive statistics for the project-level analysis, including lines of code added, and the number of commits, PRs, and PR rework submitted; for the individual-level analysis, including the number of commits, PRs, PR reviews and PR reviewed repositories. 



\begin{table}[h]
\centering
\caption{Descriptive statistics for repository level analysis.}
\label{tab:summary_stats}
\small
\begin{tabular}{l c c c c}
\toprule
Variable & Mean & Std. Dev. & Min & Max \\
\midrule
Copilot (Dummy)       & 0.301 & 0.459 & 0 & 1 \\
Code Added (log)      & 1.328 & 2.883 & 0 & 17.034 \\
Commits (log)         & 0.466 & 1.052 & 0 & 8.075 \\
PRs (log)    & 0.358 & 0.834 & 0 & 6.789 \\
PR Rework (log)       & 0.273 & 0.894 & 0 & 7.536 \\
\bottomrule
\end{tabular}
\end{table}

\begin{table}[ht]
\centering
\caption{Descriptive Statistics for individual level analysis.}
\label{tab:descriptive}
\small
\begin{tabular}{lcccc}
\hline
Variable & Mean & Std. Dev. & Min & Max \\
\hline
Copilot (Dummy) & 0.366 & 0.482 & 0 & 1 \\
Commits (log) & 1.750 & 1.694 & 0 & 10.116 \\
PRs (log) & 0.955 & 1.155 & 0 & 6.201 \\
PR Reviews (log) & 0.789 & 1.212 & 0 & 6.028 \\
PR Reviewed Repos (log) & 0.454 & 0.664 & 0 & 5.215 \\
\hline
\end{tabular}
\end{table}

\subsection{Estimation Models}

\subsubsection{Main Statistical Analysis}
\hfill

To estimate the effect of GitHub Copilot on repository / individual-level development and maintenance outcomes, we employed a DiD regression framework. This approach exploits the staggered introduction of Copilot across programming languages, comparing changes in outcomes for repositories / individuals use Copilot-supported languages (treatment group) with those use non-supported languages (control group) before and after Copilot’s technical preview. By examining within-repository / individual changes over time and contrasting them with contemporaneous changes in the control group, the DiD design isolates the impact of Copilot adoption from time-invariant repository characteristics and common temporal shocks.

The project level DiD model (repo \textit{i} month \textit{t}) is provided below:
\begin{equation}
\textit{Repository Level Effect}_{i,t} = \beta_{0} + \beta_{1}Copilot_{i,t} + \gamma_{i} + \delta_{t} +\epsilon_{i,t}
\end{equation}

The individual level DiD model (contributor \textit{i} month \textit{t}) is provided below:
 
\begin{equation}
\begin{split}
\textit{Individual Level Effect}_{i,t} = \beta_{0} + \beta_{1}Copilot_{i,t} + \beta_{2}\textit{Core Contributor}_{i} \times Copilot_{i,t} + \gamma_{i} + \delta_{t} +\epsilon_{i,t} 
\end{split}
\end{equation}

where \(y_{i,t}\) refers to the outcome measures (development and maintenance) for project / individual \(i\) in month \(t\). \(Copilot_{i,t}\) is the independent variable and a binary indicator that turns to 1 when the Copilot is released and functioned as our treatment. 
\(Core Contributor_{i}\) is the moderator and a binary indicator that turns to 1 when one individual is identified as core contributor by pretreatment code contribution behaviour. The project-level / individual-level fixed effects are represented by \(\gamma_{i}\), and \(\delta_{t}\) represents the monthly fixed effects. \(\epsilon_{i,t}\) is the robust standard error clustered at the project / individual level to account for the potential heteroskedasticity of the errors. The statistical results are presented in the appendix (Section 8.1).

\subsubsection{Parallel Trends:}
\hfill

We followed prior research (e.g.,Autor 2003) to conduct an event study and test whether PR rework before the treatment are similar and parallel, an important assumption for the DiD framework. Specifically, the equations to test the parallel
trends are:
\begin{equation}
\label{eq:eventstudy}
\text{PR rework}_{it}
= \alpha 
+ \sum_{\tau \neq -1} \beta_{\tau}
\left( Copilot_{i} \times \mathds{1}\{ t - T_i = \tau \} \right)
 + \gamma_{i} + \delta_{t} +\epsilon_{i,t} 
\end{equation}
where \({PR rework}_{it}\)  denotes the log number of PR rework actions in repository \(i\) at time \(t\); \(Copilot_{i}\) indicates repositories associated with Copilot-supported languages; and \(T_{i}\) denotes the Copilot introduction month. The coefficients \(\beta_{\tau}\) capture the relative changes in rework activity \(\tau\) months before or after treatment, using \(\tau = -1\) as the omitted baseline. Repository fixed effects (\(\gamma_{i}\)) absorb time-invariant heterogeneity, while month-year fixed effects (\(\delta_{i}\)) account for aggregate shocks.

This event-study design enables us to visually and statistically evaluate the parallel-trends assumption and to characterize how Copilot’s effects unfold over time, offering a richer interpretation than a single aggregated DiD estimate.

\section{Main Results}
\hfill

\subsection{Project Level Analysis - Technical Debt}

We analyzed the number of lines of code added, commits, and PRs as code development activities, and PR rework as a measurment of technical debt. Table \ref{results_project} presents the results. After the introduction of GitHub Copilot, repositories whose primary programming language was endorsed by Copilot (treatment) experienced a significant increase in code development activities, such as the number of lines of code added increased 17.7\% (\(\exp(0.163) - 1 \approx 17.7\%\)), commits increased 4.1\% (\(\exp(0.04) - 1 \approx 4.1\%\)), and PRs increased 4.3\% (\(\exp(0.042) - 1 \approx 4.3\%\)). This increase of productivity aligns with the analysis of industry experts \citep{Peng2023} and the findings of recent academic research \citep{yeverechyahu2024impact}. 


\renewcommand{\arraystretch}{1} 
\begin{table}[htbp]
\caption{The impact of technical preview on development and maintenance activities.}
\label{results_project}
\centering 
\begin{threeparttable} 
\begin{tabular}{l c c c c }  
 \toprule 
 Activity: & \multicolumn{3}{ c }{Development} & \multicolumn{1}{ c }{Technical Debt}  \\
 DV:
  & \multicolumn{1}{ c }{Lines of Added Code}& \multicolumn{1}{ c }{Commits}  & \multicolumn{1}{ c }{Pull Requests} & \multicolumn{1}{ c }{PR Rework}\\ \hline
Copilot & 0.163*** &     0.04*   & 0.042***  &   0.024**  \\
    &(0.06) &     (0.022)    &   (0.015)  &   (0.01) \\

\hline

Project FE   &  \checkmark  &  \checkmark  &  \checkmark  &  \checkmark   \\
Month FE    &  \checkmark  &  \checkmark  &  \checkmark &  \checkmark  \\
PR Controls    &    &    &   &  \checkmark  \\

N       &     66,120     &     66,120      &    66,168  &   66,168 \\  
Adj. $R^2$  &      0.516     &      0.603     &        0.691     &    0.81 \\
\hline \bottomrule
\end{tabular}
\begin{tablenotes}
\emph{Note:} All DVs are log-transformed. Robust standard errors clustered at the project level are presented in parentheses. $^{*}$p$<$0.1; $^{**}$p$<$0.05; $^{***}$p$<$0.01
\end{tablenotes}
\end{threeparttable}
\end{table}

More importantly, we find that the submitted PRs requires greater amount of rework, indicating a likely decrease in the quality of the submitted PRs. When analyzing the PRs submitted, the increased amount of code resubmissions is significant, even when controlling for the number of PRs, indicating an increased demand for code reviews per unit of code development. Specifically, we find that treatment repositories experienced 2.4\% (\(\exp(0.024) - 1 \approx 2.4\%\)) more code rework, keeping the number of PRs submitted constant. These AI paired contributions are less likely to adhere to repository standards and therefore generate additional code rework effort, causing the accumulation of technical debt. 

\begin{figure}[htbp]
\centering
\includegraphics[width=0.6\textwidth]{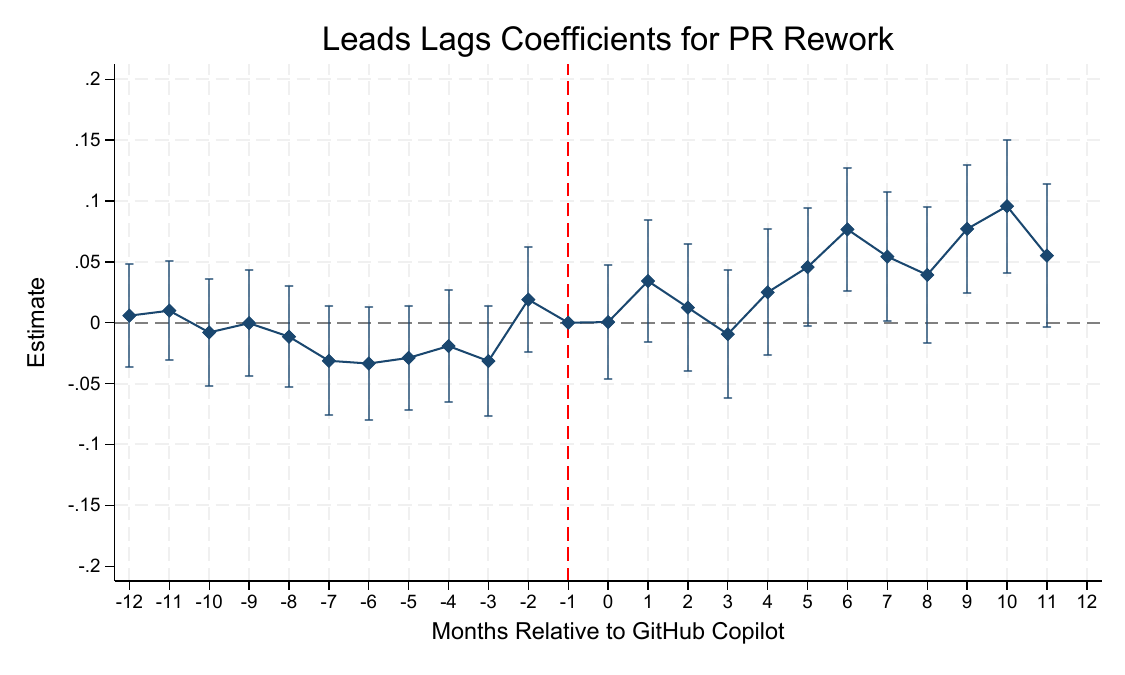}
    \caption{Parallel trends and dynamic effects of the Copilot treatment for PR rework
    }
    \label{fig:lead_lag}
\end{figure}

Figure \ref{fig:lead_lag} presents the event time analysis of the treatment effect on PR rework (leads-lags estimates). The coefficients for the pre-treatment periods are statistically insignificant, indicating that the parallel trends assumption holds \citep{angrist2009mostly}. The leads-lags estimate also indicate a long-term effect of Copilot. With the increasing deployment and adoption of AI, we expect more contributor engagement and frequent usage of Copilot. There is an upward trend in the post-treatment coefficients for PR rework that gradually becomes more positive over time (regression results are provided in the Supplementary Material). The observed pattern suggests a growing trend of increased code rework in repositories that supported AI coding partners. 

\subsection{Individual Level Analysis - Maintenance Efforts}
We next examined commits and PRs as code development activities, and PR reviews (PR controlled) and PR reviewed repositories (PR controlled) as maintenance related activities. Based on our DiD analysis (Table  \ref{results_individual_maintainer}), we find that core contributors engage in fewer development activities and in more maintenance activities after the launch of Copilot. The core contributors performed significant less code commits submissions, a decrease of 42.9\% (\(\exp(0.357) - 1 \approx 42.9\%\)) compare to the rest of contributors. At the same time, the core contributors reviewed more PRs and do so across a broader set of repositories. This shift suggests that core contributors are not only stretching more of their limited time on maintenance tasks but also spreading themselves thinner across more repositories. The resulting burden of maintenance may come at the cost of reduced productivity among more experienced contributors.


\renewcommand{\arraystretch}{1} 
\begin{table}[htbp]
\caption{The impact of technical preview on GitHub's core contributor behavior.}
\label{results_individual_maintainer}
\centering 
\begin{threeparttable} 
\begin{tabular}{l c c c c}  
 \toprule 
 Activity: & \multicolumn{2}{ c }{Development} & \multicolumn{2}{ c }{Maintenance}\\
 DV:
  & \multicolumn{1}{ c }{Commits}  & \multicolumn{1}{ c }{Pull Request} & \multicolumn{1}{ c }{PR Review}  & \multicolumn{1}{ c }{PR Reviewed Repos}  \\ \hline
Copilot & 0.142***   & 0.091***  & 0.04 &  0.008  \\
    &     (0.408)    &   (0.03) & (0.023) &  (0.012)  \\
Core Contributor  & -0.357***  & -0.027  &  0.06*  &  0.045** \\
$\times$ Copilot & (0.052) & (0.042) & (0.035) & (0.018)\\
\hline

Individual FE  &  \checkmark  &  \checkmark  &  \checkmark  &  \checkmark \\
Month FE &  \checkmark   &  \checkmark  &  \checkmark  &  \checkmark \\
PR Controlled &  \   &  \  &  \checkmark  &  \checkmark \\
N       &     51,408     &    51,408  & 51,408 & 51,408  \\  
Adj. $R^2$  & 0.643 & 0.607 &  0.799 & 0.757 \\
\hline \bottomrule
\end{tabular}
\begin{tablenotes}
\emph{Note:} All DVs are log-transformed. Robust standard errors clustered at the individual level are presented in parentheses. $^{*}$p$<$0.1; $^{**}$p$<$0.05; $^{***}$p$<$0.01
\end{tablenotes}
\end{threeparttable}
\end{table}

\section{Robustness Check}
We performed a series of tests to ensure that our findings are robust to various model specifications. Together, these tests increase confidence that our results are not driven by our model specification or unobserved variables.

\subsection{Matching Results}
As a robustness check, we employed Coarsened Exact Matching (CEM) to mitigate concerns about potential selection bias and ensure a more balanced comparison between treatment and control groups. CEM allows us to pre-process the data by matching units on a set of covariates that may influence both treatment assignment and outcomes. We matched repositories based on pretreatment code development characteristics, including Code Added, Commits, and PRs. After matching, the repository level dataset consisting of n = 2,510 repositories, with 1,486 in the treatment group and 1,024 in the control group. The treatment and control groups were more closely aligned in their baseline characteristics, reducing imbalance across key variables. We then re-estimated the DiD models using the matched sample. The results are listed in the table below. The results remain consistent with our main findings, reinforcing the conclusion that the integration of GitHub Copilot is associated with increased rework and maintenance related activities.

\begin{table}[htbp]
\centering
\caption{Balance Statistics: Unmatched and Matched Samples}
\label{tab:balance_stats}
\small
\begin{tabular}{
    l
    S[table-format=1.3] S[table-format=1.3] S[table-format=3.4]
    S[table-format=2.3] S[table-format=1.3]
    S[table-format=1.3] S[table-format=1.3] S[table-format=3.4]
    S[table-format=2.4] S[table-format=1.3]
}
\toprule
\multirow{2}{*}{Variable} & 
\multicolumn{4}{c}{Unmatched Sample} & 
\multicolumn{4}{c}{Matched Sample} \\
\cmidrule(lr){2-5} \cmidrule(lr){6-9}
& {Treated} & {Control} & {t-stat} & {p} 
& {Treated} & {Control} & {t-stat} & {p} \\
\midrule
Code Added (log)      & 3.670 & 4.143  & -2.648 & 0.008 & 3.298 & 3.519  & -1.253 & 0.210 \\
Commits (log)         & 1.470 & 1.664  &  -2.524 & 0.012 & 1.320 & 1.402 & -1.105 & 0.269 \\
Pull Request (log)    & 1.066 & 1.202 &  -2.149 & 0.032 & 1.009 & 0.974  &  -0.565 & 0.572 \\
\bottomrule
\end{tabular}
\end{table}

\renewcommand{\arraystretch}{1} 
\begin{table}[htbp]
\caption{The impact of technical preview on project development and code quality measures after CEM Matching.}
\centering 
\begin{threeparttable} 
\begin{tabular}{l c c c c }  
 \toprule 
 Concept: & \multicolumn{3}{ c }{Development} & \multicolumn{1}{ c }{Technical Debt}  \\
 DV:
  & \multicolumn{1}{ c }{Code Added}& \multicolumn{1}{ c }{Commit}  & \multicolumn{1}{ c }{Pull Request} & \multicolumn{1}{ c }{PR Rework}\\ \hline
Copilot & 0.202*** &     0.051**   & 0.038***  &   0.019**  \\
    &(0.06) &     (0.021)    &   (0.015)  &   (0.01) \\
\hline

Project FE   &  \checkmark  &  \checkmark  &  \checkmark  &  \checkmark   \\
Month FE    &  \checkmark &  \checkmark  &  \checkmark &  \checkmark  \\
PR Controls    &    &    &   &  \checkmark  \\
N       &     60,240 &     60,240      &    60,240  &   60,240 \\  
Adj. $R^2$  &      0.463  &      0.517     &        0.634     &    0.779 \\
\hline \bottomrule
\end{tabular}
\label{results_project_matching}
\begin{tablenotes}
\emph{Note:} All DVs are log-transformed. Robust standard errors clustered at the project level are presented in parentheses. $^{*}$p$<$0.1; $^{**}$p$<$0.05; $^{***}$p$<$0.01
\end{tablenotes}
\end{threeparttable}
\end{table}

\subsection{Sensitivity Analysis}

To assess the robustness of our estimated treatment effects on PR rework, we implement the coefficient stability approach proposed by \cite{Oster2019}. This method evaluates how sensitive the estimated DiD treatment effect is to potential omitted-variable bias by comparing changes in the coefficient of interest and the explanatory power of the model when moving from a baseline specification (with only time and repository fixed effects) to a fully controlled specification (including all covariates described in Section 2). Following \cite{Oster2019}, we compute the bias-adjusted treatment effect under the assumption that the degree of selection on unobservables is proportional to the degree of selection on observables, using \(\delta\) = 1 as the benchmark and setting the maximum possible \(R^2\) (R* ) at 1.3× the \(R^2\) of the fully controlled model.

For the PR rework outcome, the Oster-adjusted coefficient remains positive and statistically meaningful across all \(delta\) values greater than 1, indicating that an unrealistically large amount of omitted-variable selection would be required to attenuate the estimated Copilot effect to zero. This stability suggests that the increase in PR rework observed in our DiD estimates is unlikely to be driven by unobserved confounding and provides further confidence that our findings reflect a genuine shift in maintenance activities following the introduction of AI-assisted coding tools.

\begin{table}[t]
\centering
\caption{Sensitivity of Copilot Effect on PR Rework to Omitted Variable Bias (Oster \(\delta\)-analysis)}
\label{tab:oster}
\begin{tabular}{lcc}
\toprule
\textbf{Panel A. Regression Estimates} & \textbf{Uncontrolled} & \textbf{Controlled} \\
\midrule
Copilot coefficient        & 0.037 & 0.023 \\
R-squared                  & 0.000   & 0.818   \\
Controls included          & None    & PR, repo $\times$ time FE \\
\midrule
\textbf{Panel B. Oster Sensitivity Analysis} & \multicolumn{2}{c}{} \\
\midrule
\(\delta\) (delta) estimate                 & \multicolumn{2}{c}{\textbf{7.82}} \\
\(R_{\max}\) (assumed maximum \(R^2\))      & \multicolumn{2}{c}{1.00} \\
Implied strength of unobservables           & \multicolumn{2}{c}{7.82 $\times$ observed selection} \\
\bottomrule
\end{tabular}
\vspace{0.5em}
\end{table}

\section{Individual Level Heterogeneous Analyses}
\hfill

To examine heterogeneity among contributors, we conduct a series of subgroup analyses that exploit differences in contributors’ prior experience and roles within OSS projects. Following established practices in the OSS and information systems literature \citep{rullani2012}, we classify contributors based on their level of past commit contributions during the pretreatment period, which serves as a proxy for experience and project embeddedness. This approach allows us to distinguish between core contributors, who are typically responsible for code integration and maintenance, and peripheral contributors, who contribute less frequently and are more likely to focus on feature additions. We then estimate our DID models separately for each subgroup, enabling us to assess whether the effects of AI-assisted programming differ systematically across contributor types. 

\subsection{Sub Group Statistical Analysis}
The individual level subgroup DiD model (contributor \textit{i} month \textit{t}) is provided below:
 
\begin{equation}
\begin{split}
\textit{Subgroup Individual Level Effect}_{i,t} = \beta_{0} + \beta_{1}Copilot_{i,t} \times Sub Group_{i} + \delta_{t} +\epsilon_{i,t} 
\end{split}
\end{equation}

where \(y_{i,t}\) refers to the outcome measures (development and maintenance) for individual \(i\) in month \(t\). \(Copilot_{i,t}\) is the IV and a binary indicator that turns to 1 when the Copilot is released and functioned as our treatment. 
\(Sub Group_{i}\) is the moderator and a binary indicator that turns to 1 when one individual belongs to a subset by pretreatment code contribution behaviour. The individual-level fixed effects are represented by \(\gamma_{i}\), and \(\delta_{t}\) represents the monthly fixed effects. \(\epsilon_{i,t}\) is the robust standard error clustered at the individual level to account for the potential heteroskedasticity of the errors.

\subsection{Sub Group Analysis Results}

We calculated the quantity of commits submitted during the pretreatment period, and split the contributors accordingly into four percentile groups: 0 to 25\%, 25\% to 50\%, 50\% to 75\% and 75\% to 100\%. The results are presented in Figure \ref{fig:2x2graphs}. The statistical results are presented in the appendix (Section 8.2). 

\begin{figure}[htbp]
    \centering
        
    \begin{subfigure}{0.45\textwidth}
        \centering
        \includegraphics[width=\textwidth]{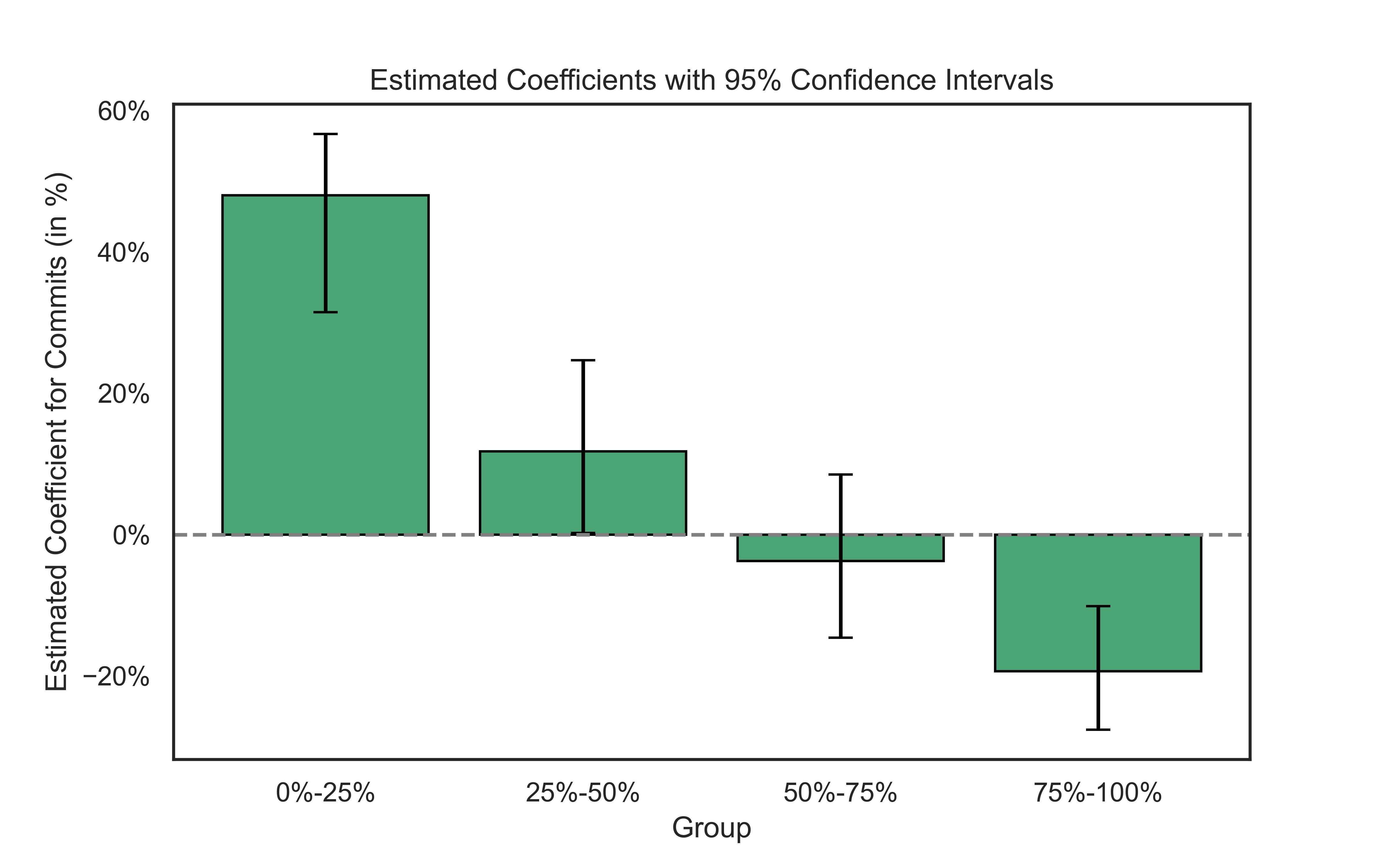}
        \caption{Commits.}
    \end{subfigure}
    \begin{subfigure}{0.45\textwidth}
        \centering
        \includegraphics[width=\textwidth]{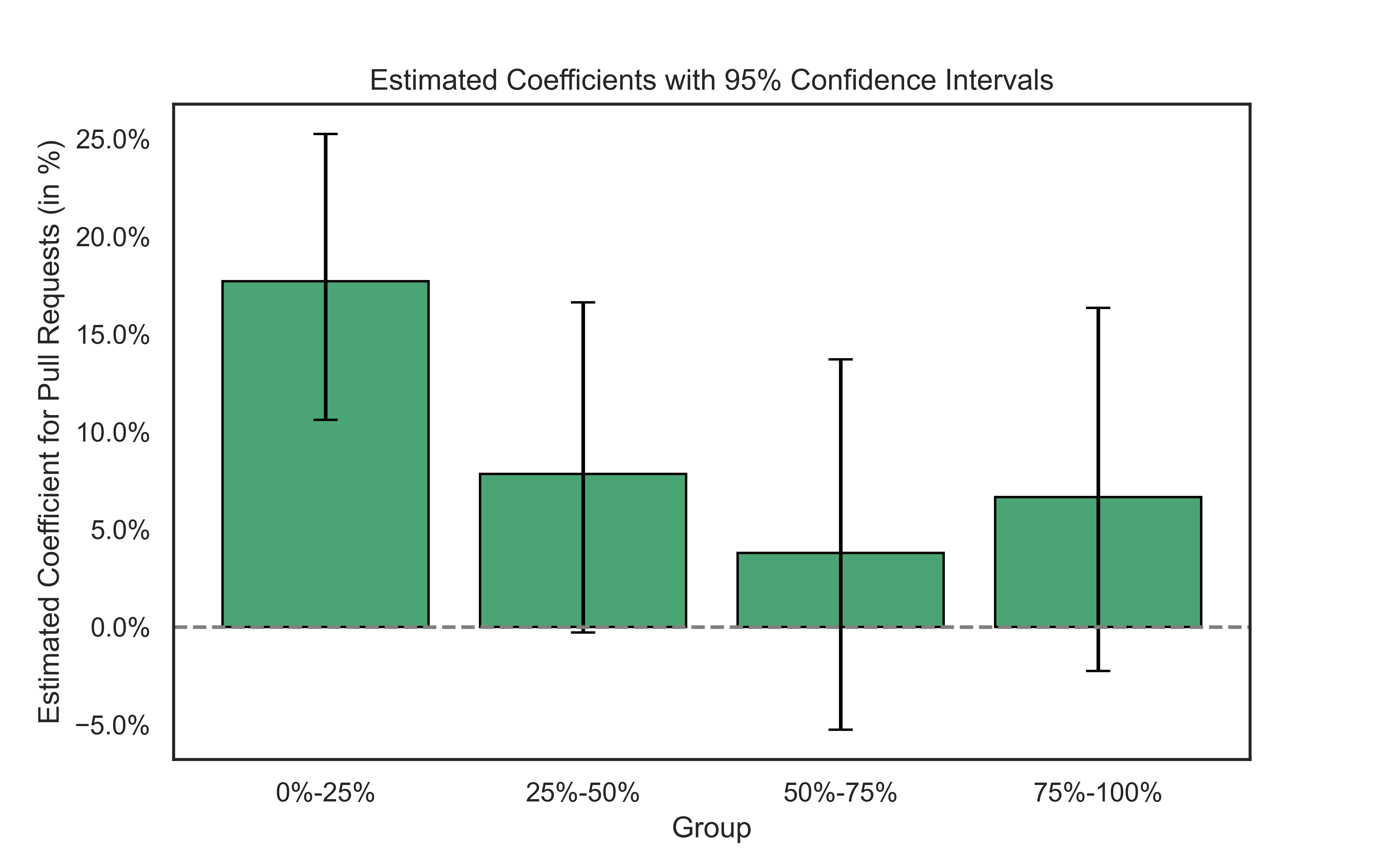}
        \caption{Pull Request.}
    \end{subfigure}
    
    \begin{subfigure}{0.45\textwidth}
        \centering
        \includegraphics[width=\textwidth]{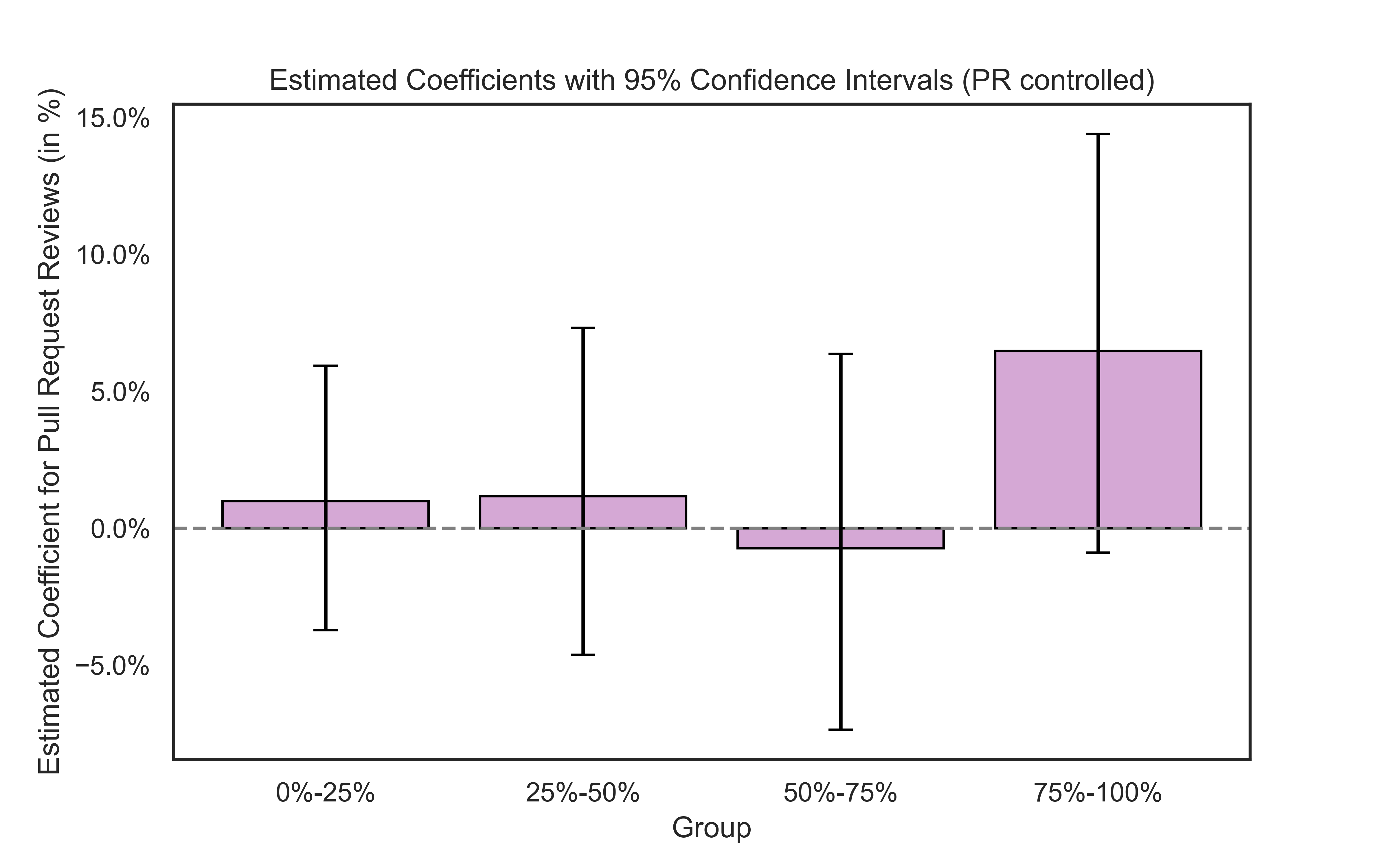}
        \caption{PR Reviews.}
    \end{subfigure}
    \begin{subfigure}{0.45\textwidth}
        \centering
        \includegraphics[width=\textwidth]{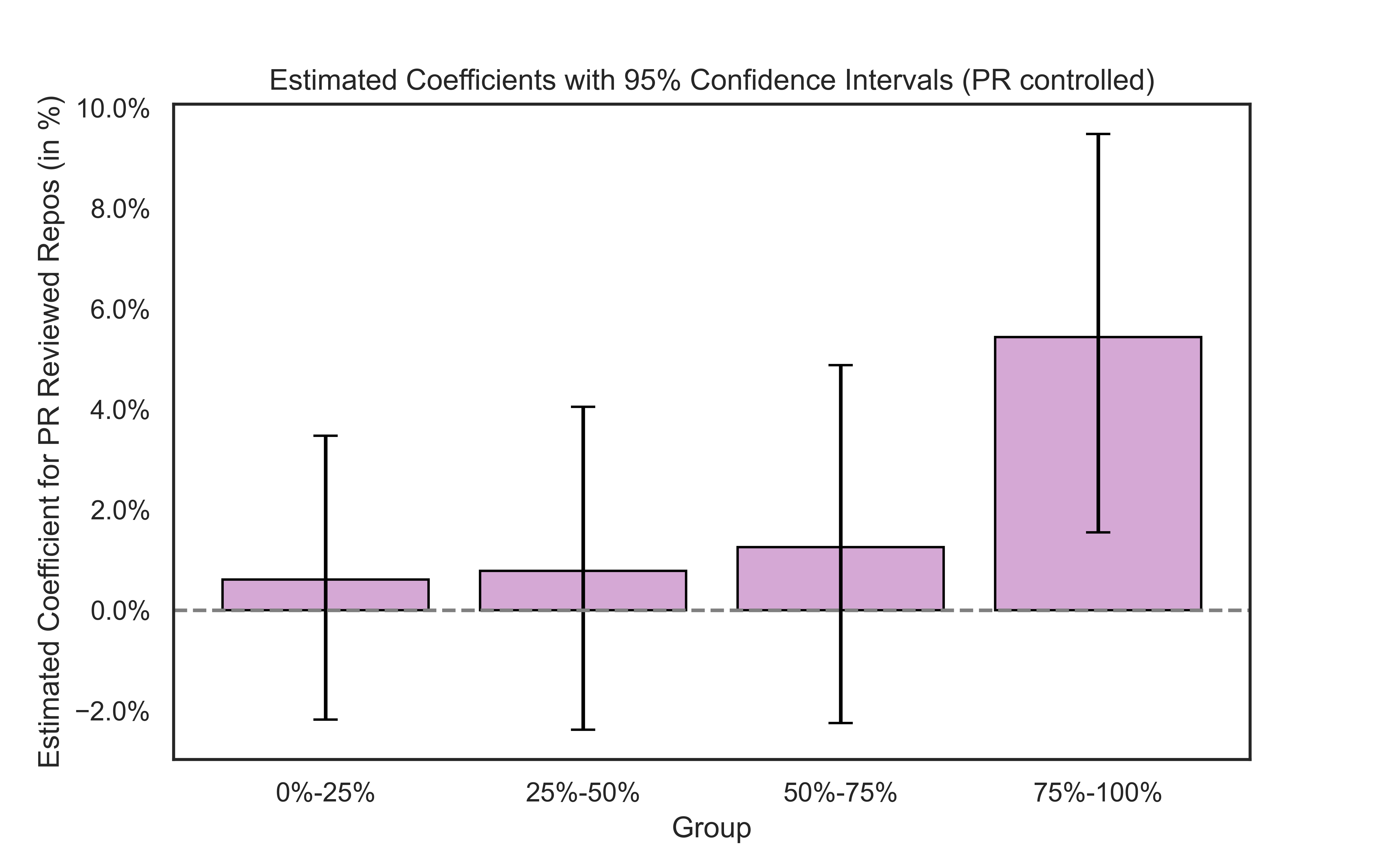}
        \caption{PR Reviewed Repos.}
    \end{subfigure}

    \caption{Contribution activities analysis by contributor subgroup. Panels show estimated coefficients (converted to \%) from DiD regressions with 95\% confidence intervals, capturing the relative change in activity post-Copilot exposure compared to control repositories. (a) Commits: Conversely, commit activity declines progressively with contributor's experience (measured by pretreatment commits quantity), with the top 25\% experiencing a 19\% reduction, suggesting reduced hands-on coding engagement. (b) Pull Requests: The most peripheral contributors (0–25\%) significantly increase their PR submissions (17.7\%), highlighting increased participation from less experienced developers. (c) PR Reviews: The top 25\% of contributors (core) exhibit a significant increase in review activity, suggesting a shift of responsibility towards the core contributors. (d) PR Reviewed Repositories: Similarly, only the core contributor group shows a meaningful rise in the number of distinct repositories reviewed, indicating a broader oversight role.}
    \label{fig:2x2graphs}
\end{figure}

Compared to the control group, we observe a consistent decline in commit volume among core contributors, while peripheral contributors show the opposite trend. Core contributors shifted towards more maintenance work, with a 19\% decrease in commits and a 6.5\% increase in PR reviews. In contrast, peripheral contributors, particularly those in the bottom percentiles, increased their commit activity by 43.5\% and submitted 17.7\% more PRs. This pattern suggests that while Copilot lowered the barriers for peripheral contributors to participate, it placed a greater maintenance burden on core contributors, redirecting their efforts from development-related to maintenance-related activities.

Our analyses show that Copilot adoption increases PR rework, indicating a rise in technical debt within OSS projects. This pattern is consistent with two complementary reasoning. First, less-experienced contributors may rely on vibe coding or shortcut-style AI-assisted development, generating larger volumes of code that require substantial revision to meet project quality standards. Second, more-experienced contributors - who traditionally provide high-quality foundational code - appear to shift their time away from original development toward reviewing and correcting others’ contributions. As a result, the supply of high-quality code creation declines while the demand for expert maintenance increases, jointly producing the observed escalation in technical debt.

\section{Discussion}
\hfill

\subsection{Key Findings and Discussion}

A growing literature documents that the adoption of GenAI technologies delivers substantial productivity gains but also introduces a range of unintended secondary effects. These include anchoring and overreliance on AI-generated suggestions \citep{Chen2024MS}, the reinforcement of societal and algorithmic biases \citep{ceciwilliamsSciencepp2025, nicoletti2023}, reduced diversity in outputs \citep{Doshi2024}, widening worker inequalities \citep{HumlumVestergaard2025}, AI–AI bias \citep{Laurito2025}, and dynamic feedback effects that amplify initial distortions over time \citep{glickman2025human}. Meta-analytic evidence further suggests that while human–AI teams excel in content generation tasks, they may underperform in judgment and coordination-intensive settings relative to humans or AI alone \citep{vaccaro2024combinations}. Our study contributes to this literature by showing how these broader secondary (and unintended) consequences of GenAI adoption manifest in software development through the lens of technical debt.

In OSS communities, the central concern is not merely whether AI increases output, but how it affects code quality and long-term maintainability. By lowering the skill threshold for writing functional code \citep{dakhel2023}, AI tools broaden participation but also increase the likelihood that contributors rely on AI-generated outputs without fully understanding their design implications or downstream consequences \citep{Barrett2023}. This dependency raises the risk that fragile, underspecified, or poorly integrated code enters production systems, thereby accelerating technical debt accumulation. Given the growing reliance of firms and public institutions on OSS infrastructure \citep{Nagle2019MS}, the economic and security consequences of such debt can be substantial.

As a case in point, in 2011 a major vulnerability (nicknamed Heartbleed) in an OpenSSL OSS project was overlooked and included in an update of the project. It went unnoticed for years, allowing any sophisticated hacker to capture secure information being passed to vulnerable web servers, including passwords, credit card information, and other sensitive data. The widely held view on Heartbleed underscores the risks of under-resourced maintenance related activities in OSS projects: :\footnote{https://mashable.com/archive/heartbleed-bug-websites-affected} 
\begin{quote}
“The mystery is not that a few overworked volunteers missed this bug; the mystery is why it hasn’t happened more often” \citep[p. 13]{eghbal2016roads}. 
\end{quote}

Empirically, we document three key findings. First, the introduction of GitHub Copilot leads to higher development activity at both the repository and individual levels, measured by commits and PR, consistent with industry evidence on AI-driven productivity gains \citep{Peng2023}. Second, these gains are accompanied by a significant increase in maintenance-related activities, as AI-generated contributions require more revisions before integration - an early indicator of technical debt accumulation. Third, the effects are highly heterogeneous: core contributors review more PRs, contribute fewer commits, and extend their maintenance responsibilities across a wider range of repositories, suggesting that AI-assisted contributions from peripheral developers increase coordination and review burdens.

By conceptualizing GenAI as an endogenous shock to software production, this study advances the technical debt literature in several important ways. First, consistent with prior work linking technical debt to firm performance \citep{Banker2021} and remediation costs \citep{Ramasubbu2016, RamasubbuKemerer2021}, we demonstrate that AI-assisted code increases realized remediation effort, measured through PR rework. Second, we show that technical debt is increasingly an outdated workload distribution phenomenon: maintenance costs are concentrated among a shrinking pool of core contributors, whose own productive output declines as maintenance demands rise. Third, our findings complement research on organizational design and autonomy \citep{ParamithaMassacci2023, YooCraigheadSamtani2025} by revealing how technological change can exacerbate asymmetries in effort and responsibility, even when the OSS repository workflow remains unchanged.

To illustrate the scale of Copilot’s impact on OSS communities, Microsoft core contributors in our dataset conduct on average, 976 commits, 160 PRs, and 166 PR reviews annually before its introduction. The increased volume of code associated with Copilot adoption results in an additional workload -- each core contributor is expected to review approximately 10 more PRs annually. This added maintenance burden corresponds to a reduction of 164 commits and 9 PR contributions per year per core contributor. More critically, GitHub’s 2024 surveys reveal that more than one-third of contributors to the 10 most popular OSS projects made their first contribution after signing up for GitHub Copilot, highlighting a significant influx of new and often less experienced developers\footnote{https://github.blog/news-insights/octoverse/octoverse-2024/}. 

With annual contributions to OSS projects approaching 1 billion, this surge in participation significantly increases the burden on core contributors, who take on the maintenance related tasks in the project. As a result, maintainers are compelled to reallocate their time toward reviewing and managing code submissions instead of writing new code.

\subsection{Contributions and Future Research}

Extant research on AI pair programming has primarily emphasized productivity and efficiency gains, suggesting that tools such as GitHub Copilot can substantially accelerate software development \citep{Peng2023}. While these benefits are evident in our data, our findings reveal a more nuanced set of consequences. In particular, we show that AI-assisted programming also amplifies software maintenance challenges, especially for core contributors who bear responsibility for code review and integration. Our individual-level analysis demonstrates that while less-experienced contributors increase their output, experienced contributors face higher review workloads and a concomitant decline in their own development activity. These results highlight a redistribution of effort within OSS communities that has received limited attention in prior work.

From a technical debt perspective, our findings suggest that AI-assisted programming alters the intertemporal trade-off between short-term development speed and long-term maintainability. The shortcuts enabled by AI tools may accelerate the output while introducing code that is difficult to integrate, extend, or refactor. The widespread use of AI-assisted pair programming - and, in extreme cases, “vibe coding” - can inject a larger volume of difficult-to-maintain code \citep{Pimenova2025, FawzyTahirBlincoe2025} into software projects, accelerating the accumulation of technical debt. Such contributions create latent liabilities for projects, as maintainers must later invest substantial effort to review, revise, or rewrite code to meet repository standards. In this sense, AI does not merely increase the volume of contributions; it changes the composition of incoming code in ways that intensify technical debt accumulation.

A key contribution of our study is to operationalize technical debt at its point of entry. We conceptualize extensive PR rework as realized technical debt: the additional modification, coordination, and revision effort required to bring submitted code up to acceptable standards. This measure complements prior work that captures technical debt through architectural metrics, defect accumulation, or long-run performance outcomes. By focusing on PR-level dynamics, we provide a micro-level view of how technical debt emerges in real time and how it is managed through ongoing maintenance effort.

Our findings also raise concerns about the learning implications of AI-assisted development. With AI providing rapid solutions, peripheral contributors may engage less deeply with underlying programming principles and best practices, resulting in code that is functional but brittle. This concern echoes evidence from other AI-augmented work settings, where less-experienced workers experience large productivity gains while more skilled workers see modest improvements and increased coordination burdens \citep{Brynjolfsson2025}. In OSS settings, these dynamics can further worsen technical debt by weakening the feedback loop between contribution and learning.

These insights point to several directions for future research. Scholars could examine how different project governance mechanisms moderate AI-induced technical debt, such as automated testing, modular architectures, or formalized review protocols. Future work may also explore heterogeneity across project types, identifying which OSS projects are most vulnerable to debt accumulation under AI-assisted development. More broadly, the dynamics documented here may extend beyond OSS to other knowledge-intensive domains where AI increases output without replacing expert judgment. As OSS components are increasingly embedded in enterprise systems and public infrastructure \citep{Nagle2019MS}, understanding how AI reshapes technical debt dynamics becomes critical not only for OSS sustainability but for the resilience of the broader digital ecosystem.

As a concluding remark, the challenges observed in OSS, such as quality concerns and increased maintenance burdens driven by productivity gains among newer and less-experienced contributors, should serve as an early warning for similar risks in other knowledge-intensive domains where AI is being promoted to boost productivity and innovation.

\newpage

\bibliographystyle{informs2014}
\bibliography{references.bib}

\clearpage
\section*{Appendix}
\hfill

\subsection*{Project Level Lead Lag Analysis}
We conducted a lead-lag analysis to examine the dynamic effects of GitHub Copilot adoption on repositories' performance over time. Using a 24-month window centered around the technical preview release: 12 months before and 12 months after, we estimated the monthly treatment effects relative to the month of launch. The coefficients of the analysis is listed below:

\begin{table}[H]
\centering
\caption{Regression Results for PR rework}
\label{tab:lead_lag_regression}
\small
\begin{tabular}{lcccccc}
\toprule
Variable & Coefficient & Std. Err. & $t$ & $P>|t|$ & 95\% CI (Lower) & 95\% CI (Upper) \\
\midrule
b12     & 0.0058 & 0.0216 & 0.27 & 0.789 & -0.0366 & 0.0482 \\
b11     & 0.0099 & 0.0207 & 0.48 & 0.632 & -0.0306 & 0.0504 \\
b10     & -0.0080 & 0.0223 & -0.36 & 0.720 & -0.0518 & 0.0358 \\
b9      & -0.0004 & 0.0222 & -0.02 & 0.987 & -0.0438 & 0.0431 \\
b8      & -0.0115 & 0.0210 & -0.55 & 0.586 & -0.0527 & 0.0298 \\
b7      & -0.0313 & 0.0228 & -1.37 & 0.170 & -0.0759 & 0.0134 \\
b6      & -0.0335 & 0.0235 & -1.43 & 0.154 & -0.0796 & 0.0125 \\
b5      & -0.0289 & 0.0218 & -1.33 & 0.184 & -0.0715 & 0.0138 \\
b4      & -0.0192 & 0.0233 & -0.82 & 0.411 & -0.0648 & 0.0265 \\
b3      & -0.0315 & 0.0232 & -1.36 & 0.174 & -0.0770 & 0.0140 \\
b2      & 0.0190 & 0.0219 & 0.87 & 0.385 & -0.0239 & 0.0619 \\
\hline
b1 & \multicolumn{6}{c}{Baseline} \\ \hline
a0      & 0.0005 & 0.0240 & 0.02 & 0.983 & -0.0466 & 0.0477 \\
a1      & 0.0342 & 0.0256 & 1.34 & 0.181 & -0.0159 & 0.0843 \\
a2      & 0.0124 & 0.0266 & 0.47 & 0.641 & -0.0398 & 0.0647 \\
a3      & -0.0094 & 0.0268 & -0.35 & 0.725 & -0.0620 & 0.0431 \\
a4      & 0.0250 & 0.0265 & 0.94 & 0.346 & -0.0269 & 0.0769 \\
a5      & 0.0456 & 0.0246 & 1.85 & 0.064 & -0.0027 & 0.0940 \\
a6      & 0.0767 & 0.0259 & 2.96 & 0.003 & 0.0260 & 0.1274 \\
a7      & 0.0543 & 0.0269 & 2.02 & 0.044 & 0.0015 & 0.1072 \\
a8      & 0.0393 & 0.0286 & 1.37 & 0.169 & -0.0167 & 0.0953 \\
a9      & 0.0770 & 0.0267 & 2.88 & 0.004 & 0.0247 & 0.1294 \\
a10     & 0.0957 & 0.0278 & 3.44 & 0.001 & 0.0412 & 0.1503 \\
a11     & 0.0551 & 0.0298 & 1.85 & 0.065 & -0.0034 & 0.1136 \\
\_cons  & 0.2416 & 0.0186 & 12.98 & 0.000 & 0.2051 & 0.2781 \\
\bottomrule
\end{tabular}
\end{table}

\subsection*{Individual Level Sub Group Analysis}

We estimate the effect of Copilot on four subgroup  of contributors based on the pretreatment contribution: 0 to 25\%, 25\% to 50\%, 50\% to 75\% and 75\% to 100\%. The analysis are conducted for each of our outcome measures: PR Reviews, PR Reviewed Repos, commits and PRs. The PR Reviews, PR Reviewed Repos have PR controlled. The statistical results are presented in the following tables.  

\begin{table}[htbp]
    \centering
    \caption{Regression Results for \texttt{PR Reviews (PR controlled)}}
    \label{tab:regression_results_PRR}
    \begin{tabular}{lcccccc}
        \toprule
        & Coefficient & Std. Err. & $t$-value & $P>|t|$ & 95\% CI (Lower) & 95\% CI (Upper) \\
        \midrule
        Subgroup 0\%-25\% &  0.0099  & 0.0243  &  0.41  & 0.685  & -0.0379  & 0.0577  \\
        Subgroup 25\%-50\% &  0.0117  & 0.0300  &  0.39  & 0.697  & -0.0472  & 0.0706  \\
        Subgroup 50\%-75\% & -0.0073  & 0.0352  & -0.21  & 0.835  & -0.0764  & 0.0617  \\
        Subgroup 75\%-100\% &  0.0628  & 0.0366  &  1.72  & 0.086  & -0.0089  & 0.1345  \\
        \bottomrule
    \end{tabular}
\end{table}

\begin{table}[htbp]
    \centering
    \caption{Regression Results for \texttt{PR Reviewed Repos (PR controlled)}}
    \label{tab:regression_results_PRRR}
    \begin{tabular}{lcccccc}
        \toprule
        & Coefficient & Std. Err. & $t$-value & $P>|t|$ & 95\% CI (Lower) & 95\% CI (Upper) \\
        \midrule
        Subgroup 0\%-25\% &  0.0061  & 0.0143  &  0.43  & 0.670  & -0.0220  & 0.0342  \\
        Subgroup 25\%-50\% &  0.0078  & 0.0163  &  0.48  & 0.630  & -0.0241  & 0.0397  \\
        Subgroup 50\%-75\% &  0.0125  & 0.0179  &  0.70  & 0.486  & -0.0227  & 0.0477  \\
        Subgroup 75\%-100\% &  0.0530  & 0.0192  &  2.76  & 0.006  &  0.0154  & 0.0906  \\
        \bottomrule
    \end{tabular}
\end{table}

\begin{table}[htbp]
    \centering
    \caption{Regression Results for \texttt{Commits}}
    \label{tab:regression_results_Commit}
    \begin{tabular}{lcccccc}
        \toprule
        & Coefficient & Std. Err. & $t$-value & $P>|t|$ & 95\% CI (Lower) & 95\% CI (Upper) \\
        \midrule
        Subgroup 0\%-25\% &  0.3614  & 0.0447  &  8.08  & 0.000  &  0.2737  & 0.4492  \\
        Subgroup 25\%-50\% &  0.1115  & 0.0555  &  2.01  & 0.045  &  0.0026  & 0.2205  \\
        Subgroup 50\%-75\% & -0.0381  & 0.0610  & -0.62  & 0.532  & -0.1578  & 0.0815  \\
        Subgroup 75\%-100\% & -0.2149  & 0.0551  & -3.90  & 0.000  & -0.3230  & -0.1068  \\
        \bottomrule
    \end{tabular}
\end{table}

\begin{table}[htbp]
    \centering
    \caption{Regression Results for \texttt{PR}}
    \label{tab:regression_results_PR}
    \begin{tabular}{lcccccc}
        \toprule
        & Coefficient & Std. Err. & $t$-value & $P>|t|$ & 95\% CI (Lower) & 95\% CI (Upper) \\
        \midrule
        Subgroup 0\%-25\% &  0.1630  & 0.0317  &  5.15  & 0.000  &  0.1009  & 0.2252  \\
        Subgroup 25\%-50\% &  0.0756  & 0.0399  &  1.89  & 0.059  & -0.0027  & 0.1539  \\
        Subgroup 50\%-75\% &  0.0373  & 0.0465  &  0.80  & 0.423  & -0.0540  & 0.1285  \\
        Subgroup 75\%-100\% &  0.0644  & 0.0444  &  1.45  & 0.147  & -0.0227  & 0.1515  \\
        \bottomrule
    \end{tabular}
\end{table}


\clearpage

\end{document}